\documentclass{webofc}
\usepackage[varg]{txfonts}   
\usepackage{media9}
\usepackage{siunitx}
\usepackage{hyperref}
\usepackage{isomath} 
\newcommand{\vect}[1]{\vectorsym{#1}}
\usepackage{slashed}
\begin{document}
\title{Dynamical fermions, centre vortices, and emergent\\ phenomena}
%
%

\author{\firstname{Derek}  \lastname{Leinweber}\inst{1} \and
        \firstname{James}  \lastname{Biddle}\inst{1} \and
        \firstname{Waseem} \lastname{Kamleh}\inst{1} \and
        \firstname{Adam}   \lastname{Virgili}\inst{1}
}

\institute{Centre for the Subatomic Structure of Matter (CSSM), 
           Department of Physics, University of Adelaide, SA, 5005, Australia
}

\abstract{The non-trivial ground-state vacuum fields of QCD form the foundation of matter.  Here we
  examine the centre vortices identified within the ground-state fields of lattice QCD.  We aim to
  understand the manner in which dynamical fermions in the QCD vacuum alter the centre-vortex
  structure.  Using modern visualisation techniques, the centre-vortex structure of pure-gauge and
  dynamical-fermion fields is quantified and compared.  We then explore the impact this modified
  structure has on measures of confinement and dynamical mass generation.  The string tension of
  the static quark potential, positivity-violation in the gluon propagator, and dynamical mass
  generation in the overlap quark propagator are of particular interest.  The impact of dynamical
  fermions is significant and provides new insights into the role of centre vortices in
  underpinning both confinement and dynamical chiral symmetry breaking in QCD.  }
\maketitle
%

\section{Introduction}

The essential, fundamentally-important, nonperturbative features of the QCD vacuum fields are the
dynamical generation of mass through chiral symmetry breaking, and the confinement of quarks.  But
what is the fundamental mechanism of QCD that underpins these phenomena?  

One of the most promising candidates is the centre vortex perspective of QCD vacuum
structure. While the ideas of a centre-vortex dominated vacuum were laid down long ago~%
\cite{tHooft:1977nqb,tHooft:1979rtg,Nielsen:1979xu}, it wasn't until 1997 when Jeff Greensite,
Manfried Faber, {\it et al.}  demonstrated that lattice QCD techniques could be used to explore the
importance of these ideas~\cite{DelDebbio:1996lih,Faber:1997rp,DelDebbio:1998luz,Bertle:1999tw,%
Faber:1999gu,Engelhardt:1999xw,Bertle:2000qv,Engelhardt:2003wm}. Indeed by the end of the
millennium, the field had attracted broad interest, with a comprehensive review in 2003~\cite{Greensite:2003bk}. 

This perspective describes the nature of the nontrivial vacuum in terms of the fundamental
centre of the gauge group.  Herein our focus is on the SU(3) gauge group where centre vortices
are characterised by the three centre phases, $\sqrt[3]{1}$.

By identifying centre vortices within the ground-state fields and then removing them, a deep
understanding of their contributions can be developed.  Removal of centre vortices from the
ground-state fields results in a loss of dynamical mass generation and restoration of chiral
symmetry~\cite{Trewartha:2015nna,Greensite:2016pfc,Trewartha:2017ive,Virgili:2021jnx}, a loss of
the string tension~\cite{Langfeld:2003ev,Bowman:2010zr,Biddle:2022zgw,Leinweber:2022dpp}, a
suppression of the infrared enhancement in the Landau-gauge gluon
propagator~\cite{Langfeld:2001cz,Bowman:2010zr,Biddle:2018dtc,Biddle:2022acd}, and the possibility
that gluons are no longer confined~\cite{Biddle:2022acd}.

One can also examine the role of the centre vortices alone.  Remarkably, centre vortices produce
both a linear static quark
potential~\cite{Langfeld:2003ev,OCais:2008kqh,Trewartha:2015ida,Biddle:2022zgw,Leinweber:2022dpp}
and infrared enhancement in the Landau-gauge gluon propagator~\cite{Biddle:2018dtc,Biddle:2022acd}.
The planar vortex density of centre-vortex degrees of freedom scales with the lattice spacing
providing a well defined continuum limit~\cite{Langfeld:2003ev}.  These results elucidate strong
connections between centre vortices and confinement.

A connection between centre vortices and instantons has been identified through gauge-field
smoothing~\cite{Trewartha:2015ida}.  An understanding of the phenomena linking these degrees
of freedom is illustrated in Ref.~\cite{Biddle:2019gke}.  In addition, centre vortices have been
shown to give rise to mass splitting in the low-lying hadron spectrum~%
\cite{Trewartha:2017ive,Trewartha:2015nna,OMalley:2011aa}.

Still, the picture in pure SU(3) gauge theory is not perfect. The vortex-only string tension
obtained from pure Yang-Mills lattice studies has been consistently shown to be about $\sim 60\%$
of the full string tension. Moreover, upon removal of centre vortices the gluon propagator shows a
remnant of infrared enhancement~\cite{Biddle:2018dtc}.  In short, within the pure gauge sector, the
removal of long-distance non-perturbative effects via centre-vortex removal is not perfect.

Understanding the impact of dynamical fermions on the centre-vortex structure of QCD ground-state
fields is a contemporary focus of the centre-vortex field~%
\cite{Biddle:2022zgw,Biddle:2020isk,Leinweber:2021keq,Leinweber:2022dpp,Biddle:2022acd,Virgili:2021jnx}.
Herein, changes in the microscopic structure of the vortex fields associated with the inclusion of
dynamical fermions are illustrated. The introduction of dynamical fermions brings the phenomenology
of centre vortices much closer to a perfect encapsulation of the salient features of QCD,
confinement and dynamical mass generation through chiral symmetry breaking.

As such, it is interesting to ask, what do these centre-vortex structures look like? To this end,
we present visualisations of centre vortices as identified on lattice gauge-field configurations.
Some of these visualisations are presented as stereoscopic images.

\section{Centre Vortex Identification}

Centre vortices are identified through a gauge fixing procedure designed to bring the lattice link
variables as close as possible to the identity matrix multiplied by a phase equal to one of the
three cube-roots of 1. Here, the original Monte-Carlo generated configurations are considered.
They are gauge transformed directly to Maximal Centre Gauge~\cite{DelDebbio:1996mh,Langfeld:1997jx,Langfeld:2003ev}.  This brings the lattice link variables
$U_\mu(x)$ close to the centre elements of SU(3)
\begin{equation}
Z = \exp \left ( \frac{2 \pi i}{3}\, n \right ) \, \mathbf{I} \, , 
\label{CentreSU3}
\end{equation}
with $n = -1$, $0$, or $1$ enumerating the three cube roots of 1 such that the special property of SU(3) matrices, $\det(Z) = 1$, is satisfied.
One considers gauge transformations $\Omega$ such that,
\begin{equation}
\sum_{x,\mu} \,  \left | \mathrm{tr}\, U_\mu^\Omega(x) \, \right |^2 \stackrel{\Omega}{\to}
\mathrm{max} \, ,
\label{GaugeTrans}
\end{equation}
and then projects the link variables to the centre
\begin{equation}
U_\mu(x) \to Z_\mu(x) \textrm{ where }
Z_\mu(x) = \exp \left ( \frac{2 \pi i}{3}\, n_\mu(x) \right
)\mathbf{I} \, .
\label{UtoZ}
\end{equation}
Here, $n$ has been promoted to a field, $n_\mu(x)$, taking a value of $-1$, $0$, or $1$ for each
link variable on the lattice.
In this way, the gluon field, $U_\mu(x)$, is characterised by the most fundamental aspect of the
SU(3) link variable, the centre, $Z_\mu(x)$.  In the projection step, eight degrees of freedom are
reduced to one of the three centre phases.
This ``vortex-only'' field, $Z_\mu(x)$, can be examined to learn the extent to which centre
vortices alone capture the essence of nonperturbative QCD.

The product of these centre-projected links, $Z_\mu(x)$, around an elementary $1\times 1$ square
(plaquette) on the lattice also produces a centre element of SU(3).  The value describes the centre
charge associated with that plaquette
\begin{equation}
z = \prod_\Box Z_\mu(x) = \exp \left ( 2 \pi i\, \frac{m}{3} \right ) ,\; 
m = -1,0, \mbox{\,or\,}1\,.
\label{CentreCharge}
\end{equation}
The most common value observed has $m=0$ indicating that no centre charge pierces the plaquette.
However, values of $m = \pm 1$ indicate that the centre line of an extended three-dimensional
vortex pierces that plaquette.

\begin{figure}[t]
  \centering
  \includegraphics[width=10.0truecm]{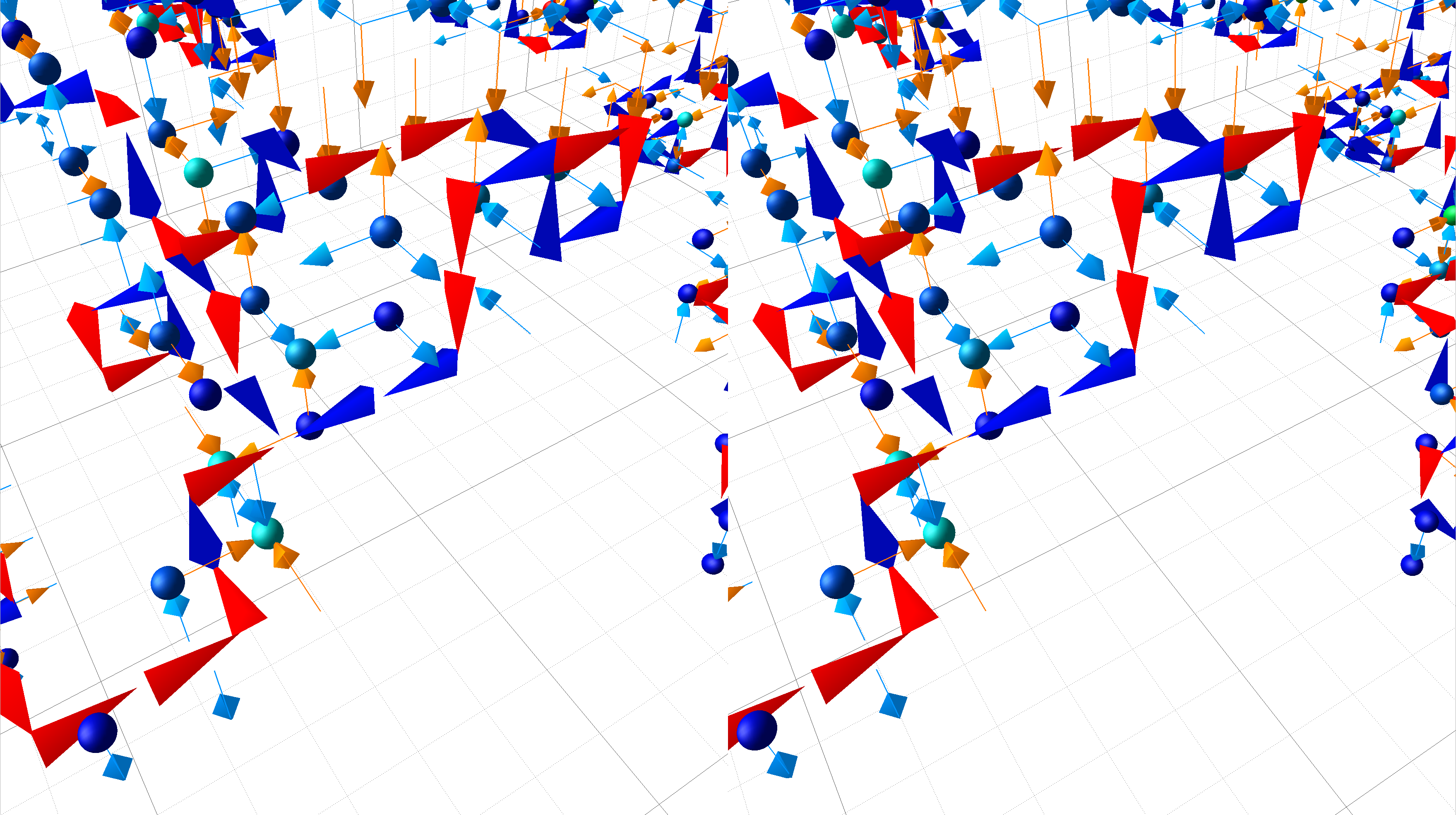}
  \caption{Stereoscopic image of centre vortices as identified on the
    lattice from Ref.~\cite{Biddle:2020isk}. Vortex features including vortex lines (jets), branching
    points (3-jet combinations), crossing points (4 jets), indicator
    links (arrows) and singular points (spheres) are described in the
    text.}
  \label{fig:VortexFeatures}
\end{figure}

The complete centre-line of an extended vortex is identified by tracing the presence of
nontrivial centre charge, $m= \pm 1$, through the spatial lattice. 
Figure~\ref{fig:VortexFeatures} exhibits rich emergent structure in the nonperturbative QCD
ground-state fields in a stereoscopic image.  Here a 3D slice of the 4D space-time lattice is being
considered at fixed time.  
To experience figure~\ref{fig:VortexFeatures} in 3D, try the
following:
\begin{enumerate}
\setlength\itemsep{-3pt}
\item If you are viewing the image on a monitor, ensure the image width is 12 to 13 cm.
\item Bring your eyes very close to one of the image pairs.
\item Close your eyes and relax.
\item Open your eyes and allow the (blurry) images to line up. Tilting your head from side to side
  will move the images vertically. 
\item Move back slowly until your eyes are able to focus.  There's no need to cross your eyes!
\end{enumerate}

\noindent
Features of the vortex phenomena include:

\paragraph{Vortex Lines:} 
The plaquettes with nontrivial centre charge, characterised by $m=+1$ or $-1$, are plotted as jets
piercing the centre of the plaquette.  Both the orientation and colour of the jets reflect the value
of the non-trivial centre charge.  Using a right-hand rule for the direction, plaquettes with
$m=+1$ are illustrated by blue jets in the forward direction, and plaquettes with $m=-1$ are
illustrated by red jets in the backward direction.  Thus, the jets show the directed flow of $m=+1$
centre charge, $z=e^{2\pi i/3}$, through spatial plaquettes.  They are analogous to the line
running down the centre of a vortex in a fluid.\\[-6pt]

\noindent
Vortices are somewhat correlated with the positions of significant topological charge density, but
not in a strong manner~\cite{Biddle:2019gke}.  However, the percolation of vortex structure is
significant and the removal of these vortices destroys most instanton-like objects.

\paragraph{Branching Points or Monopoles:} 
In SU(3) gauge theory, three vortex lines can merge into or
emerge from a single point.  Their prevalence is surprising, as is their correlation with
topological charge density~\cite{Biddle:2019gke}.

\paragraph{Vortex Sheet Indicator Links:}
As the vortex line moves through time, it creates
a vortex sheet in 4D spacetime.  This movement is illustrated by arrows along the links of
the lattice (shown as cyan and orange arrows in figure~\ref{fig:VortexFeatures}) indicating centre
charge flowing through space-time plaquettes in the suppressed time direction.

\paragraph{Singular Points:} 
When the vortex sheet spans all four space-time dimensions, it can generate
topological charge. Lattice sites with this property are called singular points~%
\cite{Bruckmann:2003yd,Engelhardt:2010ft,Engelhardt:2000wc,Engelhardt:1999xw} and are illustrated
by spheres. The sphere colour indicates the number of times the sheet adjacent to a point can
generate a topological charge contribution~\cite{Biddle:2019gke}.

\section{Impact of Dynamical Fermions}

First results demonstrating the impact of dynamical fermions on the centre-vortex structure of QCD
ground-state fields were presented in Ref.~\cite{Leinweber:2021keq}. Here we introduce improved
visualisation algorithms and explore alternative time slices.

Matched lattices are considered, one in pure-gauge and the other a $2+1$-flavor dynamical-fermion
lattice from the PACS-CS Collaboration~\cite{Aoki:2008sm}.  These $32^3 \times 64$ lattice
ensembles employ a renormalisation-group improved Iwasaki gauge action and non-perturbatively
${\cal O}(a)$-improved Wilson quarks, with $C_{\rm SW} = 1.715$.

The lightest $u$- and $d$-quark-mass ensemble identified by a pion mass of 156
MeV~\cite{Aoki:2008sm} is presented here.  The scale is set using the Sommer parameter with $r_0 =
0.4921$ fm providing a lattice spacing of $a=0.0933$ fm~\cite{Aoki:2008sm}.  A matched $32^3 \times
64$ pure-gauge ensemble using the same improved Iwasaki gauge action with a Sommer-scale spacing of
$a = 0.100$ fm is created to enable comparisons with the PACS-CS ensembles.

\begin{figure}[p]
\centering
\null\hspace{-0.3cm}
\includemedia[
        noplaybutton,
	3Dtoolbar,
	3Dmenu,
	3Dviews=U3D/Primary-Secondary-PG-24.vws,
	3Dcoo  = 16 16 32, 
        3Dc2c=0.245114266872406 0.8673180341720581 0.43321868777275085,
	3Droo  = 110.0,    
	3Droll =-98.1,     
	3Dlights=Default,  
	width=\textwidth,  
]{\includegraphics{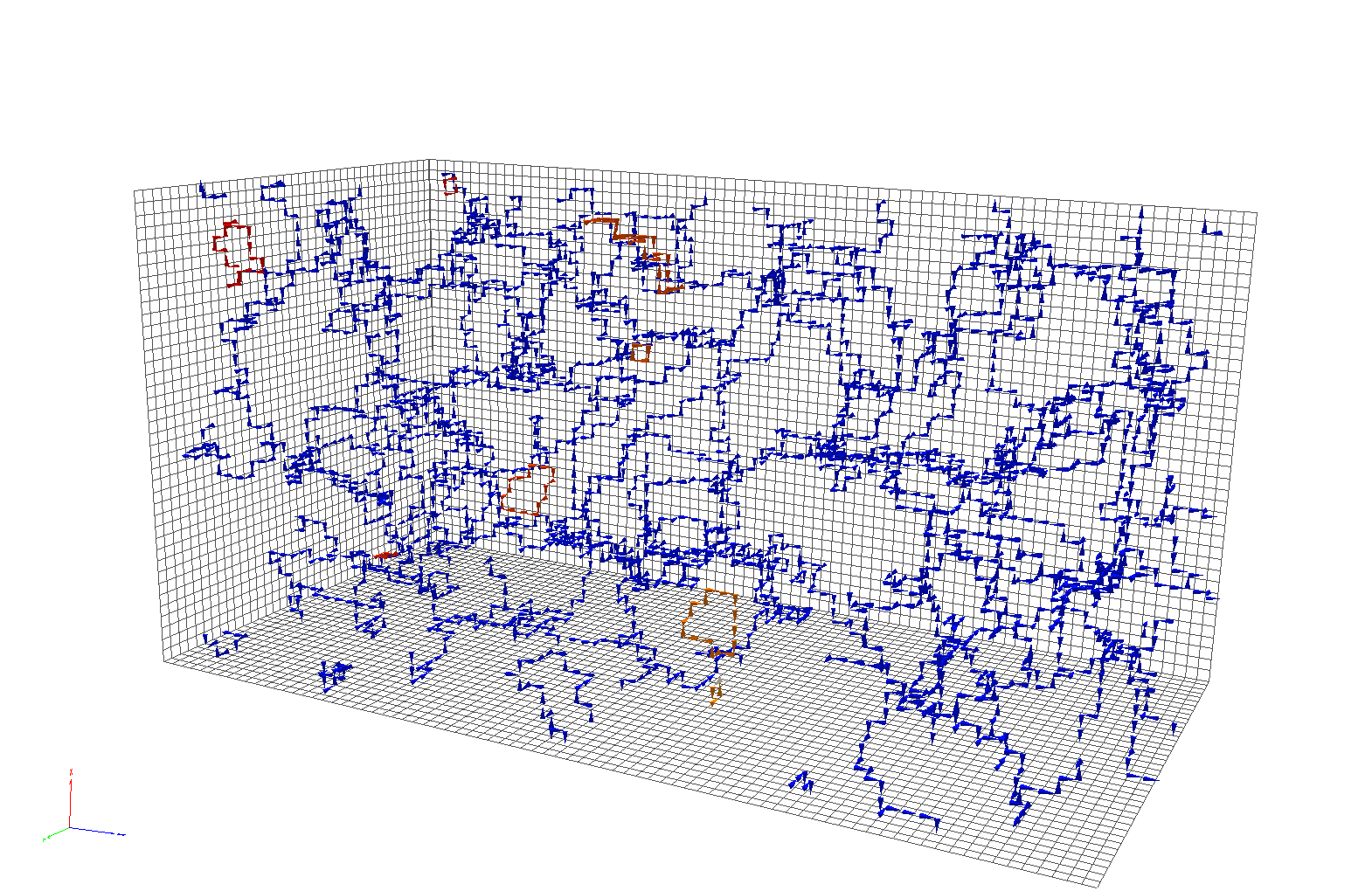}}{U3D/Primary-Secondary-PG-24.u3d}%
\caption{ The centre vortex structure of a ground-state vacuum field configuration in pure SU(3)
  gauge theory. The flow of $+1$ centre charge through the gauge field is illustrated by the jets
  (see main text for a description of the plotting conventions). Here, blue jets are used to illustrate
  the primary percolating vortex cluster, while other colours illustrate the secondary
  clusters. ({\sl Click to activate.})
\label{fig:Primary-Secondary-PG-24.u3d} }

  \centering
  \includemedia[
        noplaybutton,
	3Dtoolbar,
	3Dmenu,
	3Dviews=U3D/Primary-Secondary-DF-05.vws,
	3Dcoo  = 16 16 32, 
        3Dc2c=0.245114266872406 0.8673180341720581 0.43321868777275085,
	3Droo  = 110.0,    
	3Droll =-98.1,     
	3Dlights=Default,  
	width=\textwidth,  
]{\includegraphics{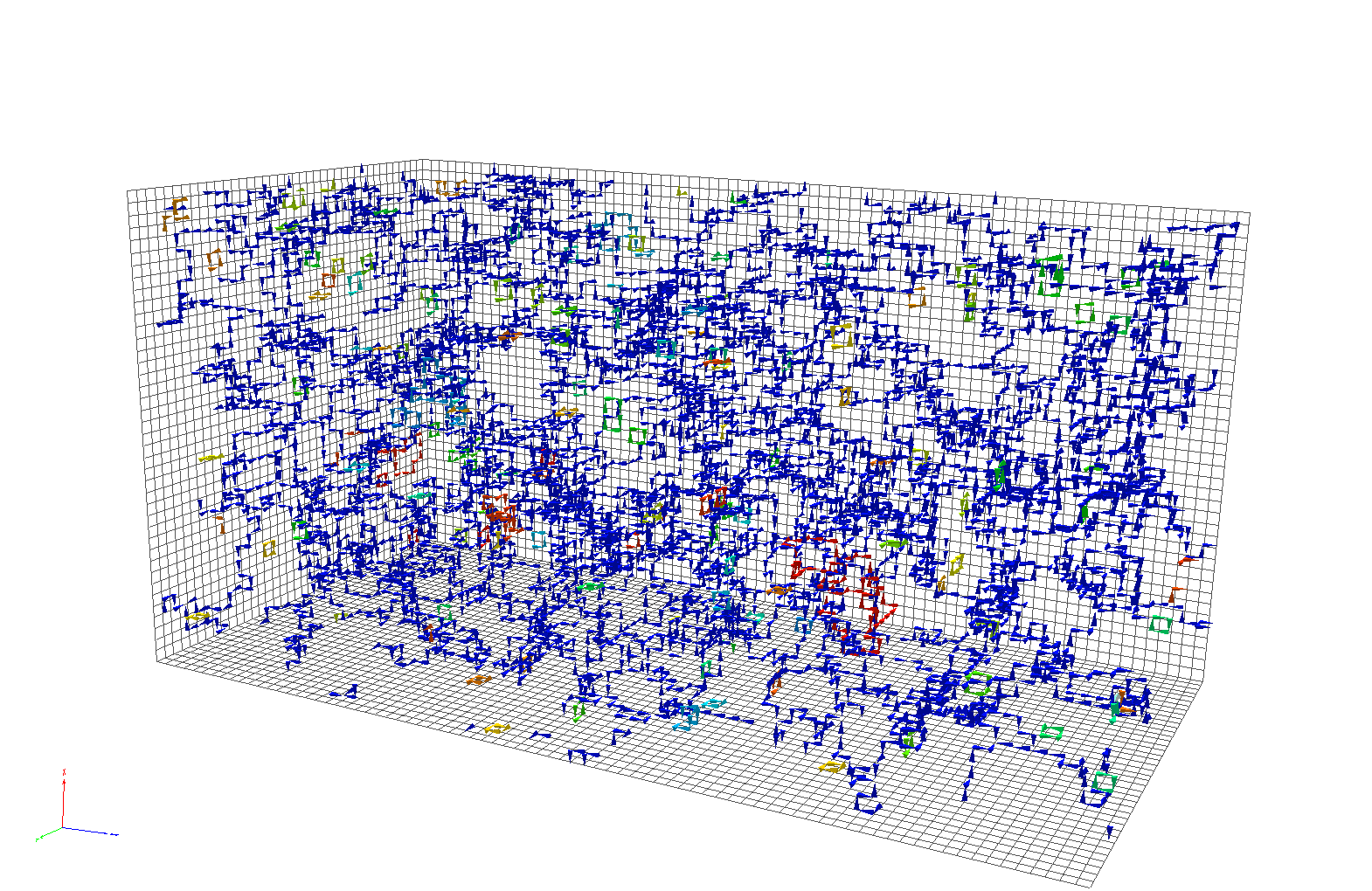}}{U3D/Primary-Secondary-DF-05.u3d}
\caption{ The centre-vortex structure of a ground-state vacuum field configuration in dynamical 2+1
  flavour QCD with $m_{\pi} = 156 ~ \si{MeV}$.  Symbols are as described in
  figure~\ref{fig:Primary-Secondary-PG-24.u3d}. ({\sl Click to activate.})
\label{fig:Primary-Secondary-DF-05.u3d}}
\end{figure}

The centre-vortex structure of pure-gauge and dynamical-fermion ground-state vacuum fields is
illustrated in figures~\ref{fig:Primary-Secondary-PG-24.u3d} and
\ref{fig:Primary-Secondary-DF-05.u3d} respectively.  These are interactive plots which can be
activated by clicking on the image in Adobe Reader\footnote{To interact with these models, it is
  necessary to open this document in Adobe Reader or Adobe Acrobat (requires version 9 or
  newer). Linux users may install
  \href{ftp://ftp.adobe.com/pub/adobe/reader/unix/9.x/9.4.1/enu/}{Adobe Acroread version 9.4.1}
  (the last edition to have full 3D support), or use a Windows emulator such as
  \href{https://www.playonlinux.com/en/}{PlayOnLinux}. From the Adobe ``Edit'' menu, select
  ``Preferences...'' and ensure ``3D \& Multimedia'' is enabled and ``Enable double-sided
  rendering'' is selected.}.  Once activated, click and drag to rotate, Ctrl-click to translate,
Shift-click or mouse wheel to zoom, and right click to access the ``Views'' menu.  Several views
identifying interesting features have been created to facilitate an inspection of the centre-vortex
structure.

\begin{figure}[p]
\centering
\includemedia[
        noplaybutton,
	3Dtoolbar,
	3Dmenu,
	3Dviews=U3D/Secondary-PG-24.vws,
	3Dcoo  = 16 16 32, 
        3Dc2c=0.245114266872406 0.8673180341720581 0.43321868777275085,
	3Droo  = 110.0,    
	3Droll =-98.1,     
	3Dlights=Default,  
	width=\textwidth,  
]{\includegraphics{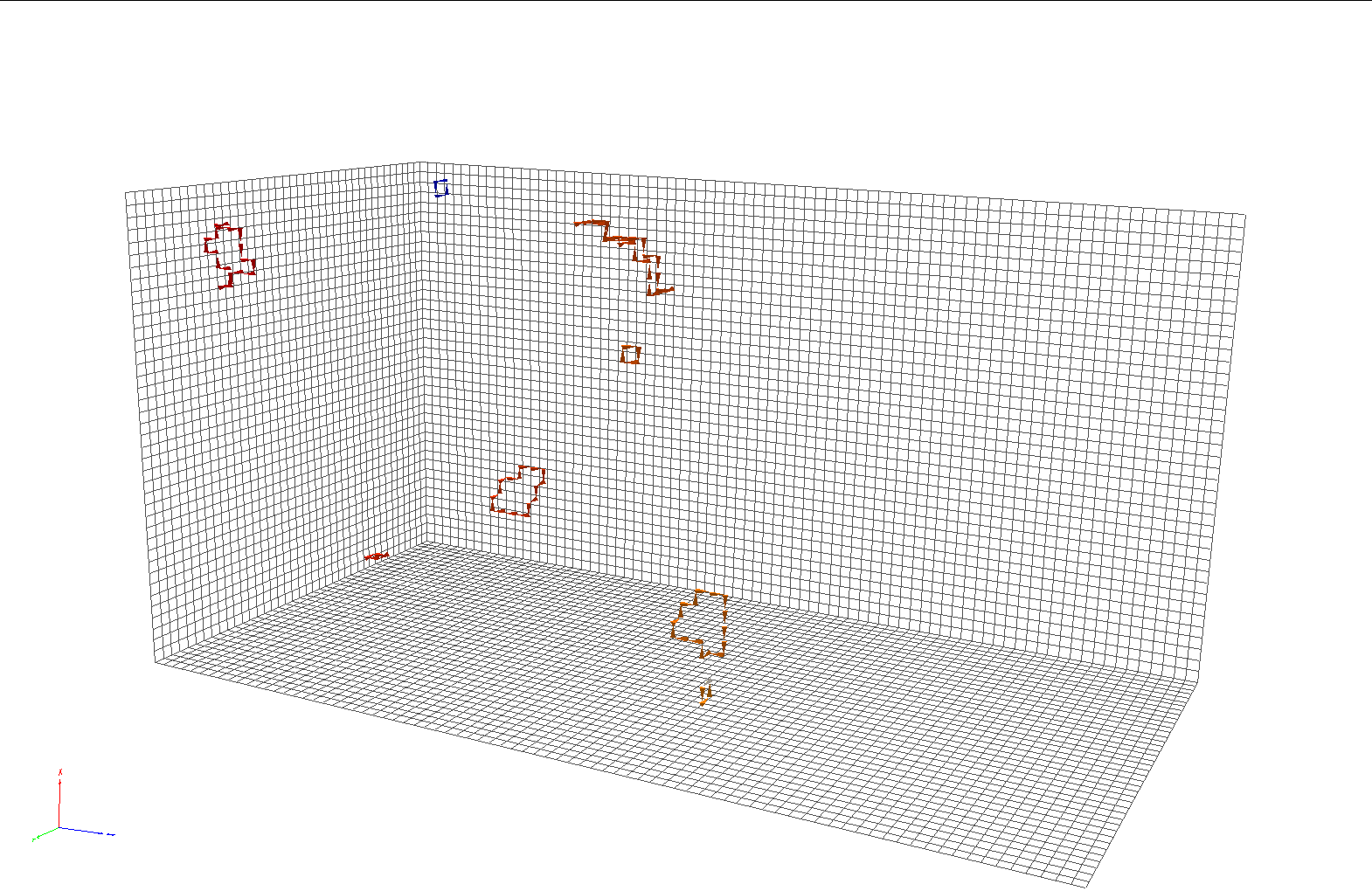}}{U3D/Secondary-PG-24.u3d}
\caption{ The centre-vortex structure of the secondary loops identified from the pure-gauge
  configuration shown in figure~\ref{fig:Primary-Secondary-PG-24.u3d}. ({\sl Click to activate.})
\label{fig:Secondary-PG-24.u3d} }

  \centering
\includemedia[
        noplaybutton,
	3Dtoolbar,
	3Dmenu,
	3Dviews=U3D/Secondary-DF-05.vws,
	3Dcoo  = 16 16 32, 
        3Dc2c=0.245114266872406 0.8673180341720581 0.43321868777275085,
	3Droo  = 110.0,    
	3Droll =-98.1,     
	3Dlights=Default,  
	width=\textwidth,  
]{\includegraphics{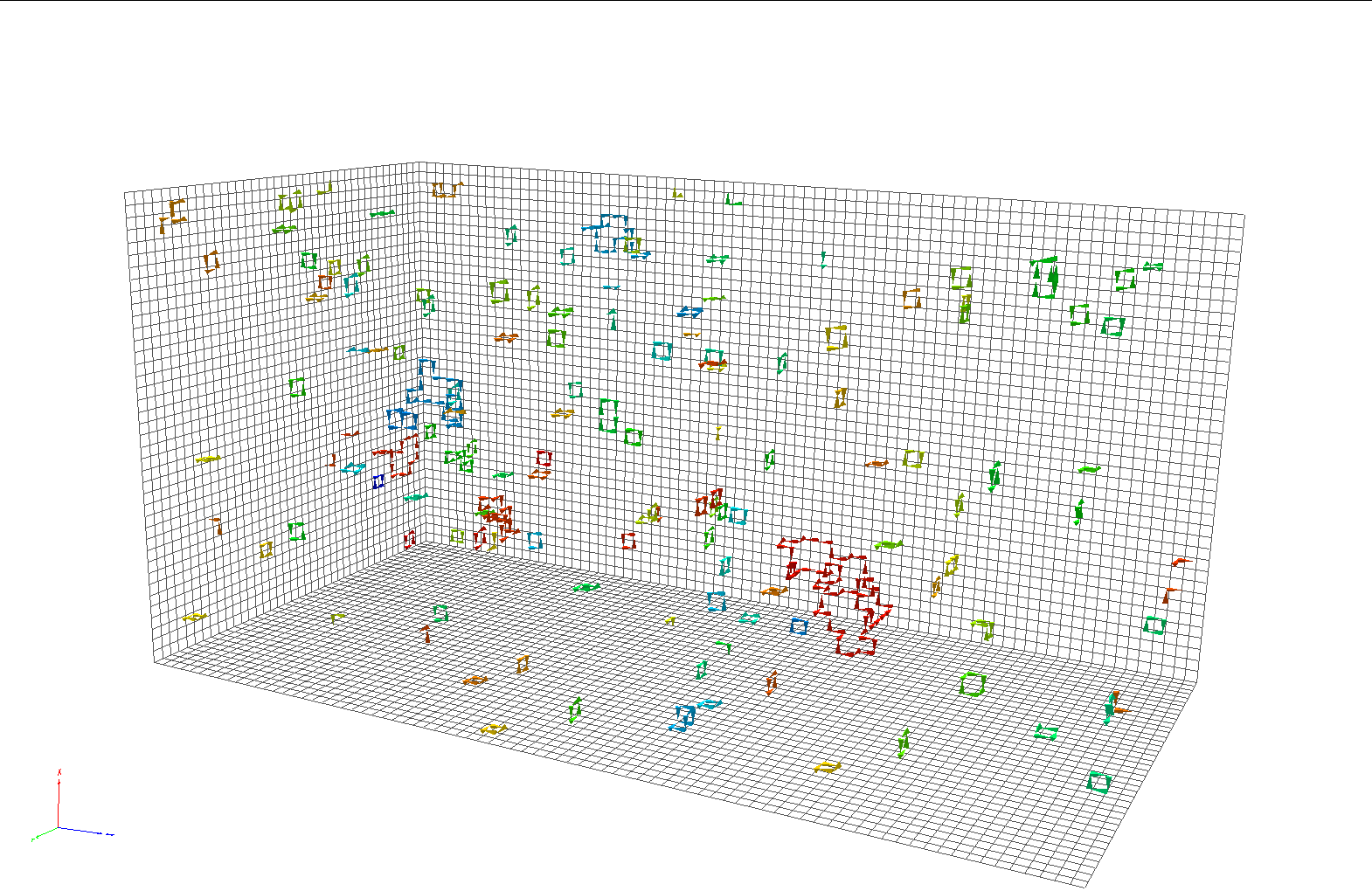}}{U3D/Secondary-DF-05.u3d}
\caption{ The centre-vortex structure of the secondary loops identified from the dynamical-fermion
  configuration shown in figure~\ref{fig:Primary-Secondary-DF-05.u3d}. ({\sl Click to activate.})
\label{fig:Secondary-DF-05.u3d}}
\end{figure}

In both illustrations, the vortex structure is dominated by a single large percolating structure, a
characteristic feature of the confining phase~\cite{Engelhardt:1999fd}.  Whereas small loops will
tend to pierce a Wilson loop twice with zero effect, it is this extended structure that gives rise
to a net vortex piercing of a Wilson loop and the generation of an area law associated with
confinement.  These two illustrations are representative of the ensemble in that the vortex
structure is typically dominated by a single large percolating cluster.

Closer inspection reveals a continuous flow of centre charge, often emerging or converging to
monopole or anti-monopole vertices where three jets emerge from or converge to a point.  These are
also referred to as branching points~\cite{Spengler:2018dxt}, as a $+1$ centre charge flowing out
of a vertex is equivalent to $+2$ centre charge flowing into the vertex and subsequently branching
to two $+1$ jets flowing out of the vertex.

Figures \ref{fig:Secondary-PG-24.u3d} and \ref{fig:Secondary-DF-05.u3d} illustrate the same time
slices as in figures ~\ref{fig:Primary-Secondary-PG-24.u3d} and
\ref{fig:Primary-Secondary-DF-05.u3d} but with the primary percolating clusters removed.  The
vortex structure of the secondary clusters in the dynamical-fermion case is more complex, typically
featuring branching points in their structure.  Again, views have been created in the interactive
figures to identify interesting features and link their locations between the full and secondary
structure illustrations.

With the introduction of dynamical fermions, the structure becomes more complex, both in the
abundance of vortices and branching points.  The average number of vortices composing the primary
cluster in these $32^2 \times 64$ spatial slices roughly doubles from $\sim 3,000$ vortices in the
pure gauge theory to $\sim 6,000$ in full QCD.  Still, there are $32^2 \times 64 \times 3 =
196,608$ spatial plaquettes on these lattices and thus the presence of a vortex is a relatively
rare occurrence.  

By counting the number of vortices between branching points one discovers the distribution is
exponential, indicating a constant branching probability.  This probability is higher in full QCD by
a ratio of $\sim 3/2$.

\section{Static Quark Potential}

With an understanding of the impact of dynamical-fermion degrees of freedom on the centre-vortex
structure of ground-state vacuum fields, attention has turned to understanding the impact on
confinement.
In a variational analysis of standard Wilson loops composed of several spatially-smeared sources to
isolate the ground state potential, the static quark potential has been calculated on three
ensembles including the original untouched links, $U_\mu(x)$, the vortex-only links, $Z_\mu(x)$,
and vortex-removed links, $Z_\mu^\dagger(x)\, U_\mu(x)$~\cite{Biddle:2022acd} where the
multiplication of the conjugate of the centre-projected field ensure all plaquettes have $z=0$.

\begin{figure}[tb]
  \centering 
  \includegraphics[width=0.49\linewidth]{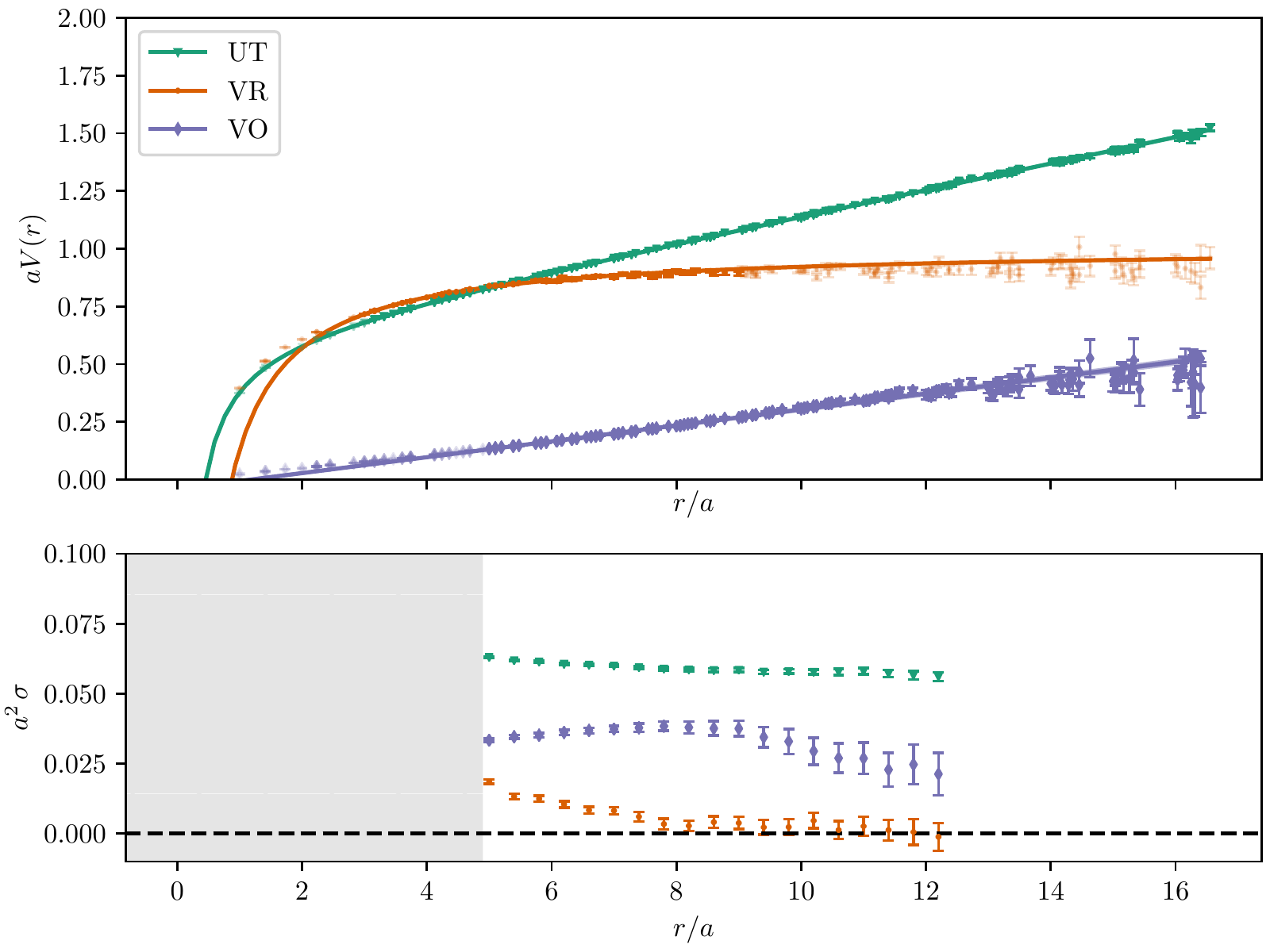} 
  \includegraphics[width=0.49\linewidth]{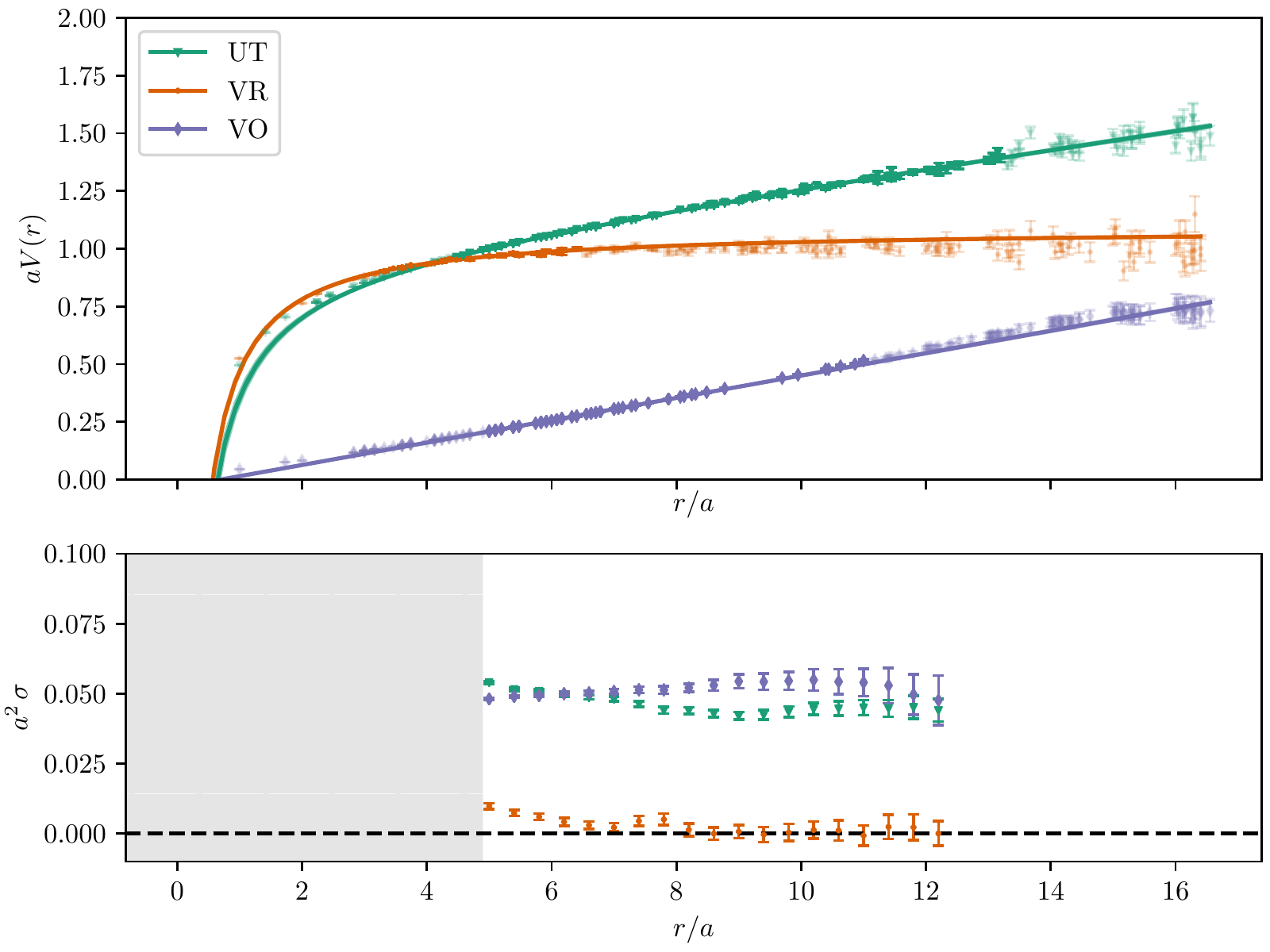} 
  \caption{The static quark potential calculated on the vortex-modified pure-gauge (left) and
    $m_\pi = 156$ MeV dynamical-fermion (right) ensembles.  The lower plot shows the local slope
    from linear fits of the potentials in the upper plot over a forward-looking window from $r
    \mbox{ to } r + 4a$.}
\label{fig:sqp}
\end{figure}

For the original untouched configurations, the static quark potential is expected to follow a
Cornell potential
\begin{equation}
  \label{eq:2}
  V(r) = V_0 - \frac{\alpha}{r} + \sigma\, r \, .
\end{equation}
As centre vortices are anticipated to encapsulate the non-perturbative long-range physics, the
vortex-only results should give rise to a linearly rising potential.  On the other hand, the
vortex-removed results are expected to capture the short-range Coulomb behaviour.
Figure~\ref{fig:sqp} from Ref.~\cite{Biddle:2022acd} illustrates the static quark potentials
obtained from these three ensembles for the pure-gauge and dynamical $2+1$-flavor ensemble with a
pion mass of 156 MeV~\cite{Aoki:2008sm}.

Qualitatively, centre vortices account for the long-distance physics.  The removal of centre
vortices completely removes the confinement potential.  And while the vortex-only string tension is
typically $~60$ \% of the original string tension in the pure gauge sector, the introduction of
dynamical fermions has improved the vortex-only phenomenology significantly.  Vortices alone
capture both the screening of the pure-gauge string tension and the full string tension of the
original untouched ensemble.
This result is associated with the significant modification of the centre-vortex structure
of ground-state vacuum fields induced by dynamical fermions.

The vortex-removed potential enables an examination of the Coulomb term in the ansatz of
Eq.~(\ref{eq:2}) and its ability to characterise the short-distance aspects of QCD in the absence
of confinement.  We find the long-range aspect of the Coulomb term to be inconsistent with the
vortex-removed potential.  This is illustrated in figure~\ref{fig:sqp} by fading out the lattice
QCD results that are inconsistent with the Coulomb ansatz.  Potentials including (anti-)screening
considerations to capture the correct physics are recommended in Ref.~\cite{Biddle:2022zgw}.
Indeed, the standard extraction of the string tension is spoiled by the persistent nature of the
Coulomb term and alternative forms are discussed in detail~\cite{Biddle:2022zgw}.

\section{Gluon propagator and positivity violation}

\begin{figure}
  \centering
  \includegraphics[width=0.48\linewidth]{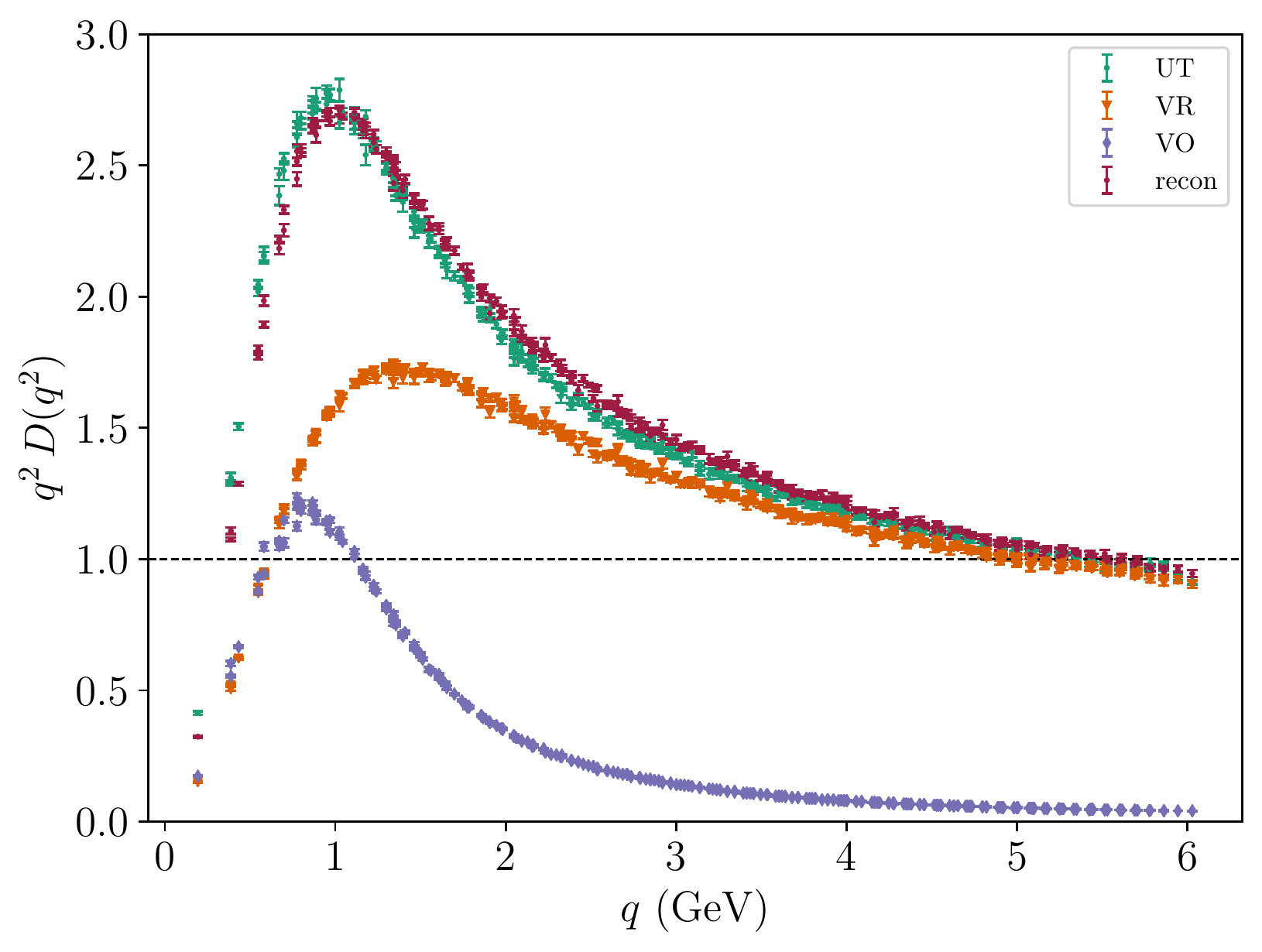} \quad
  \includegraphics[width=0.48\linewidth]{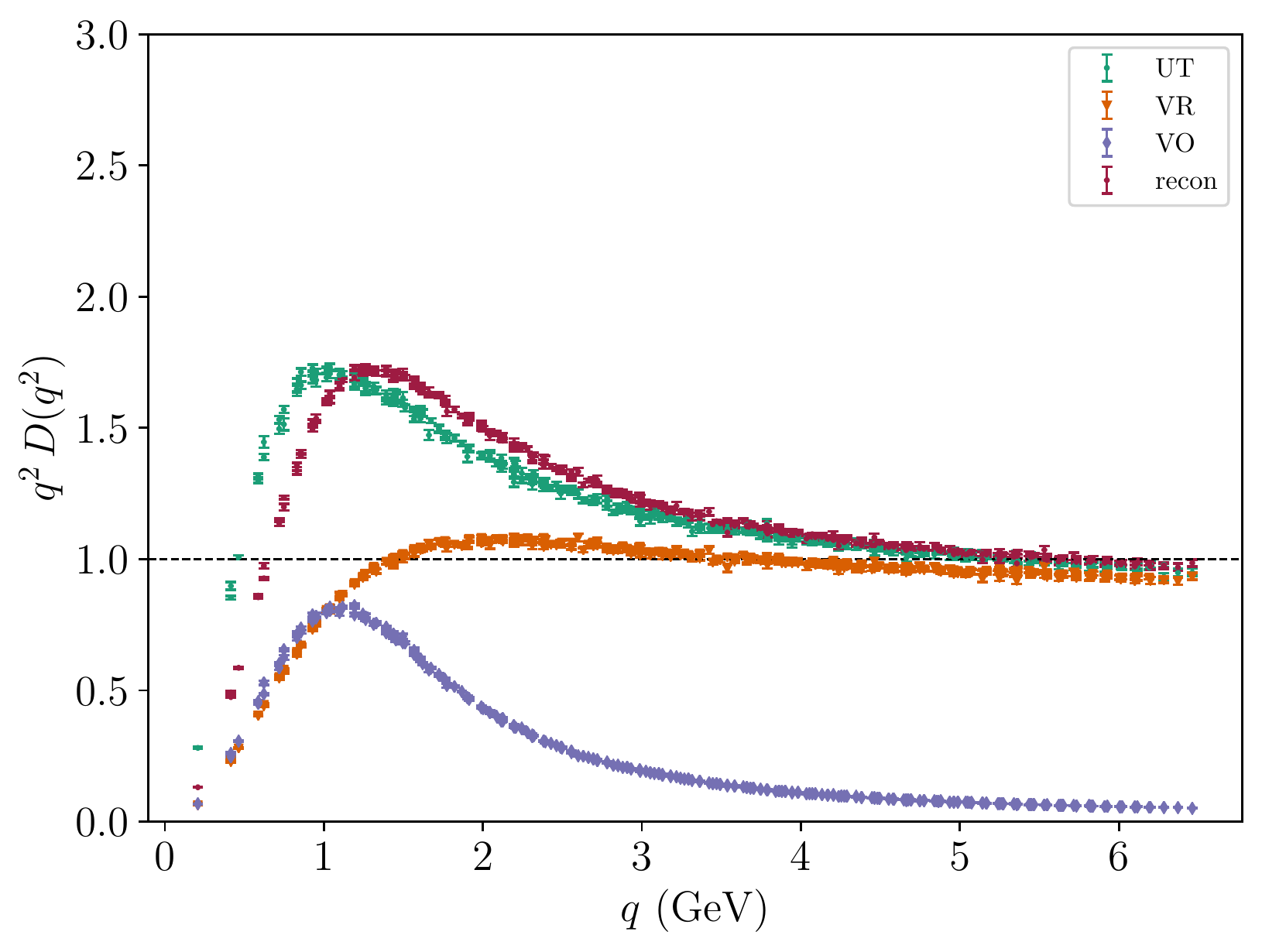}
  \caption{The pure-gauge gluon propagator (left) is compared with the $m_{\pi}=156~\si{MeV}$
    dynamical-fermion gluon propagator (right).  The propagators are renormalised as described in
    the text.  UT refers to the original untouched configurations, VR to the vortex-removed
    configurations and VO to the vortex-only configurations.  The ``recon'' results are a
    reconstruction of the original propagator obtained from a linear combination of the vortex-only and vortex-removed
    propagators.}
  \label{fig:gluon_prop}
\end{figure}

The improved separation of perturbative and nonperturbative physics through the consideration of
vortex-removed and vortex-only ensembles in full QCD is also manifest in the nonperturbative gluon
propagator~\cite{Biddle:2022acd}.  This time vortex removal removes the infrared enhancement of the
gluon propagator, leaving a tree-level like structure as illustrated in figure~\ref{fig:gluon_prop}
for the gluon-propagator renormalisation function $Z(q^2) = q^2\, D(q^2)$.  Of course,
finite-volume effects force $Z(q^2) \to 0$ as $q^2 \to 0$ as $D(q^2)$ is finite.  Details of the
renormalisation procedure are summarised at the end of Sec.~III A in Ref.~\cite{Biddle:2022acd}.

To explore positivity violation, one considers the Euclidean correlator, $C(t)$, obtained by taking
the Fourier transform of $D(q_0,\vect{0})$ such that~\cite{Biddle:2022acd}
\begin{equation}
  \label{eq:1}
  C(t)= \frac{1}{2\pi}\int_{-\infty}^{\infty}dq_0\, \int_0^{\infty}dm^2\, \frac{\rho(m^2)}{q_0^2+m^2}\,e^{-i\,q_0\,t}\, .
\end{equation}
Extending the $q_0$ integral to the complex plane and employing the residue theorem, one arrives at
\begin{equation}
  \label{eq:3}
  C(t)=\int_0^{\infty}dm\, e^{-mt}\,\rho(m^2)\,.
\end{equation}
Clearly if $C(t)<0$ for any $t$ then $\rho(m^2)$ is not positive definite, and we say that
positivity has been violated. This implies that there is no K{\"a}llen-Lehmann representation and
as such the propagator does not represent a correlation between physical states.  Hence, the states
do not appear in the physical spectrum. In the context of the gluon propagator, this can be taken
as an indication that gluons are confined.

\begin{figure}
  \centering
  \includegraphics[width=0.48\linewidth]{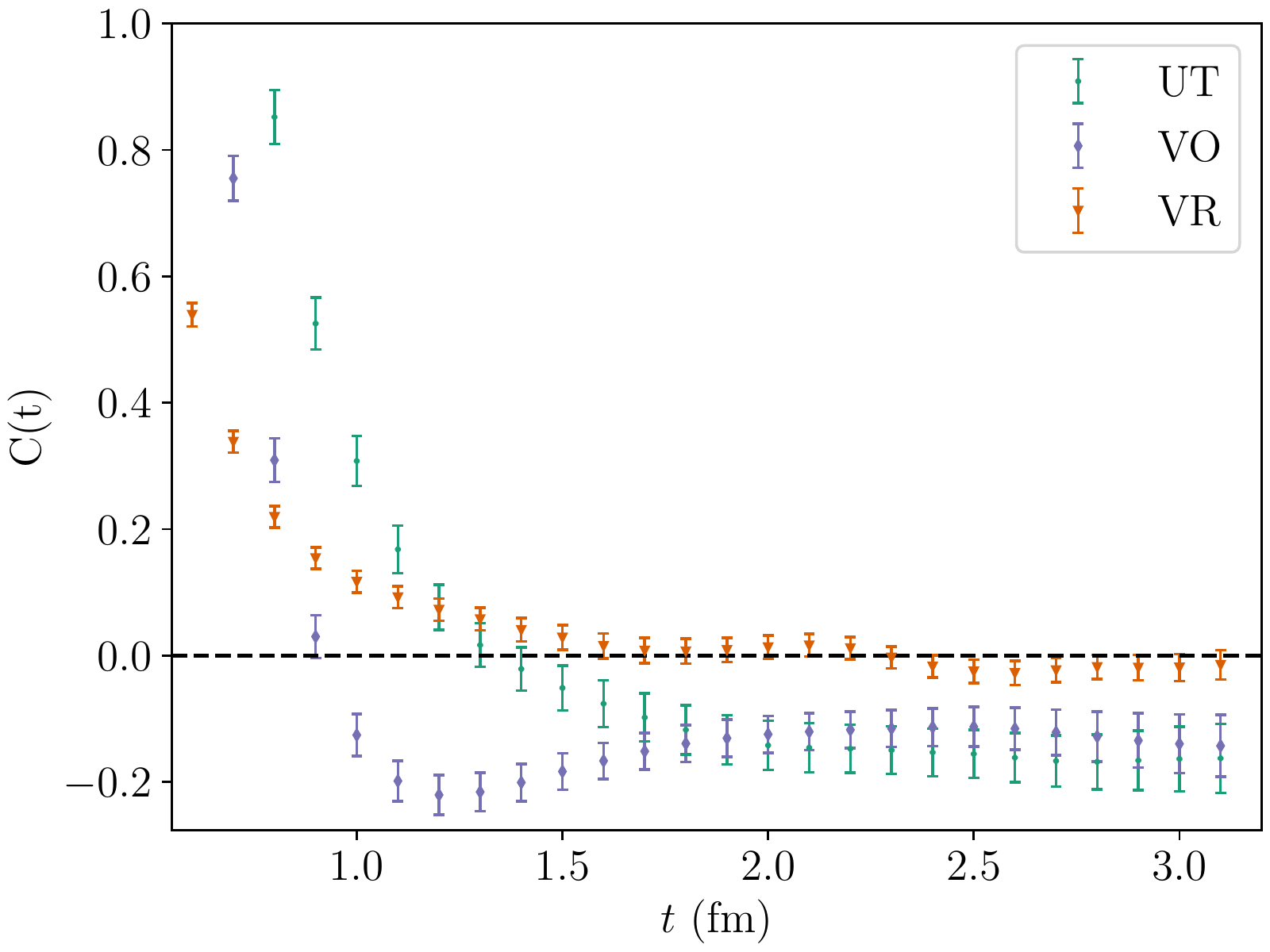} \quad
  \includegraphics[width=0.48\linewidth]{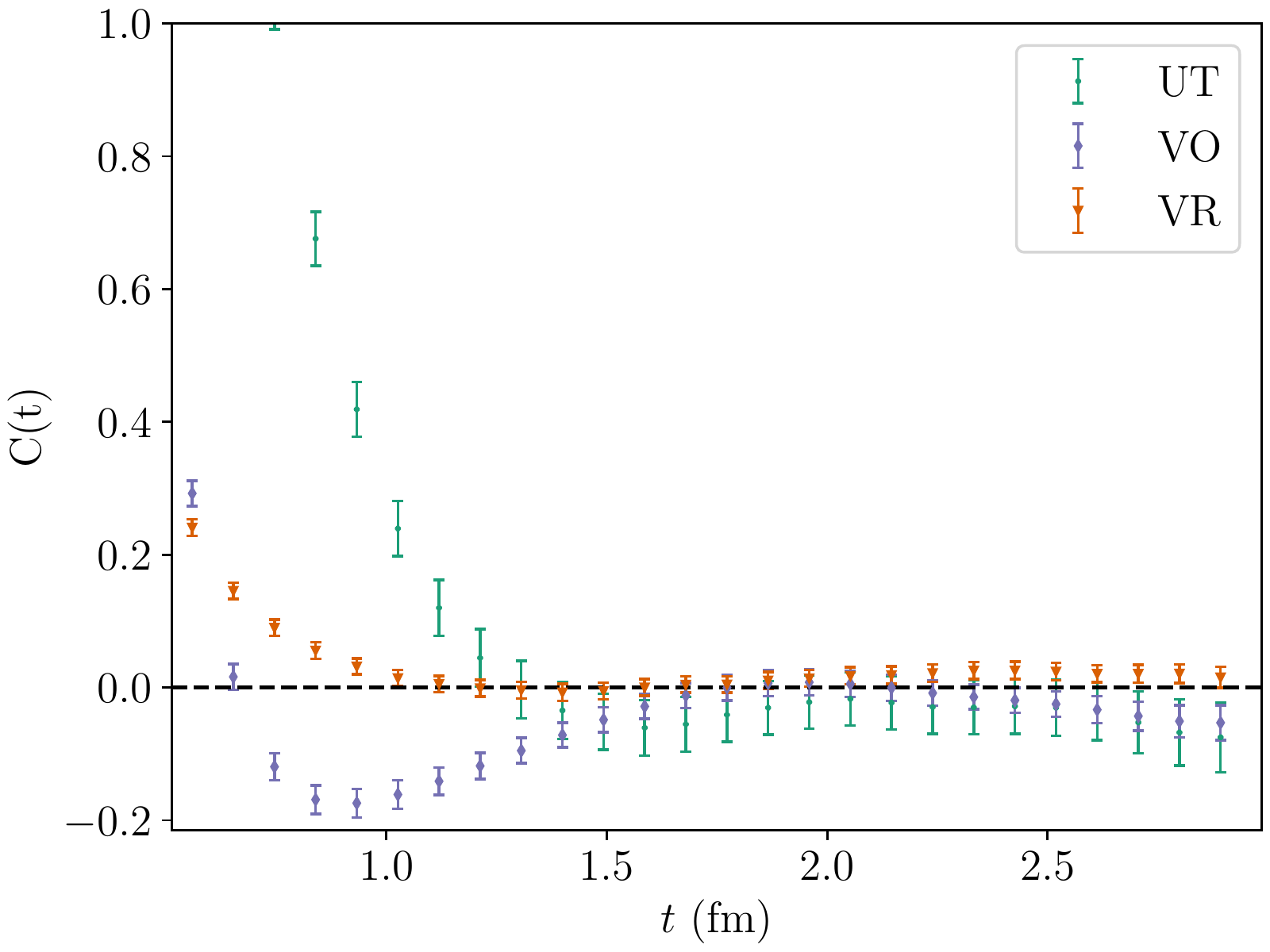}
  \caption{The pure-gauge Euclidean correlator (left) is compared with the $m_{\pi}=156~\si{MeV}$
    dynamical-fermion Euclidean correlator (right). Shown are the results from the untouched
    (UT), vortex-removed (VR) and vortex-only (VO) ensembles. A dashed line at $C(t)=0$
    is provided to aid in observing positivity violation.}
  \label{fig:EuclCorr}
\end{figure}

The numerical calculation of Eq.~(\ref{eq:3}) is described in detail in Ref.~\cite{Biddle:2022acd},
where the subtleties of the finite volume are fully accounted for.  The results are shown in
figure~\ref{fig:EuclCorr}. As expected~\cite{Bowman:2007du}, the untouched correlator shows clear signs
of positivity violation.  Interestingly, the vortex-only correlators also exhibit robust positivity
violation. 

The positivity violation present in the pure-gauge vortex-removed result at large distances is
consistent with the observations made in figure~\ref{fig:gluon_prop}, where residual infrared
strength in the vortex-removed gluon propagator is apparent. Thus, the separation of perturbative
and non-perturbative physics through vortex modification is imperfect in the pure-gauge sector.

The results from the dynamical ensemble in the right-hand plot of figure~\ref{fig:EuclCorr}
demonstrate an interesting change in behaviour.  As with the gluon propagator results in the
previous section, the most striking change is in the vortex-removed correlator. In this sector we
now observe consistency with positivity in the dynamical-fermion case. This supports the
interpretation of the positivity violation in the vortex-removed pure-gauge results as being
related to the residual non-perturbative infrared strength in the gluon propagator. As this
residual strength is significantly diminished on the dynamical ensembles, we now see that the
residual $q^2$ dependence in the VR renormalisation function may be purely perturbative in
origin. In this case, vortex modification has been successful in separating perturbative and
non-perturbative physics.

In summary, the vortex-only ensembles capture the infrared enhancement of the gluon propagator and
the screening of this enhancement in full QCD.  Moreover, the vortex-only configurations exhibit
significant positivity violation, as would be expected of a confining infrared-dominated theory.
Conversely, the vortex-removed dynamical-fermion configurations show a loss of this positivity
violation, admitting the possibility that they do support a spectral representation of the
propagator constructed from perturbative gluon interactions. These results provide additional
support for the fact that centre vortices encapsulate the confining aspects of QCD.

\section{Quark propagator and dynamical mass generation} 

The overlap Landau-gauge quark propagator $S(p)$ can be written in the form
\begin{equation}
    S(p) = \frac{Z(p)}{i\slashed{q} + M(p)}\, ,
    \label{eq:qp}
\end{equation}
where $Z(p)$ is the renormalisation function and $M(p)$ is the mass function, and $q$ is a
tree-level improved momentum variable minimising lattice artefacts~\cite{Virgili:2022wfx}.  The
infrared behaviour of the mass function, specifically, the presence of dynamical mass generation,
provides a clear signal of dynamical chiral symmetry breaking.

\begin{figure}[t]
    \centering
    \includegraphics[width=0.49\linewidth]{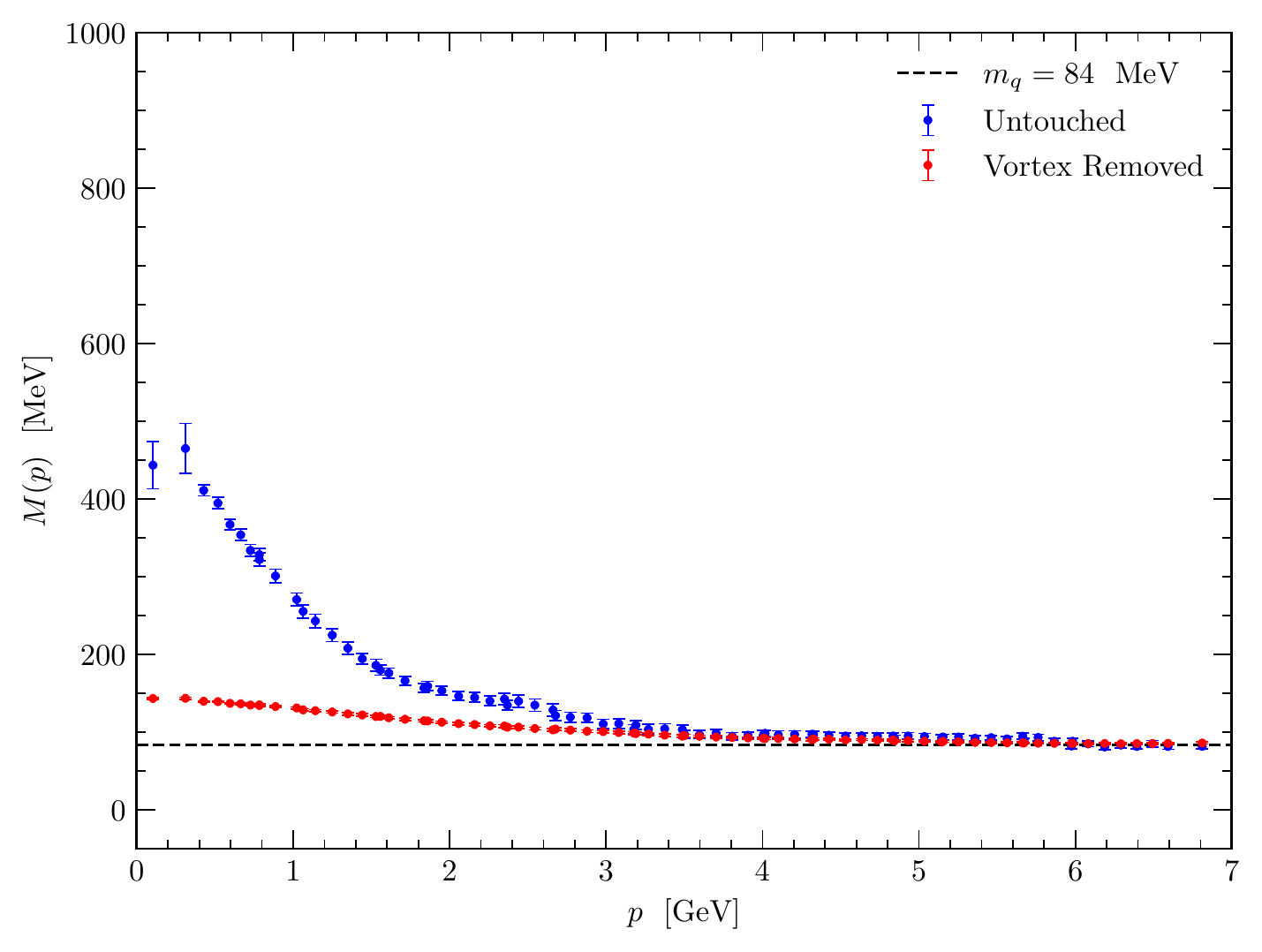} 
    \includegraphics[width=0.49\linewidth]{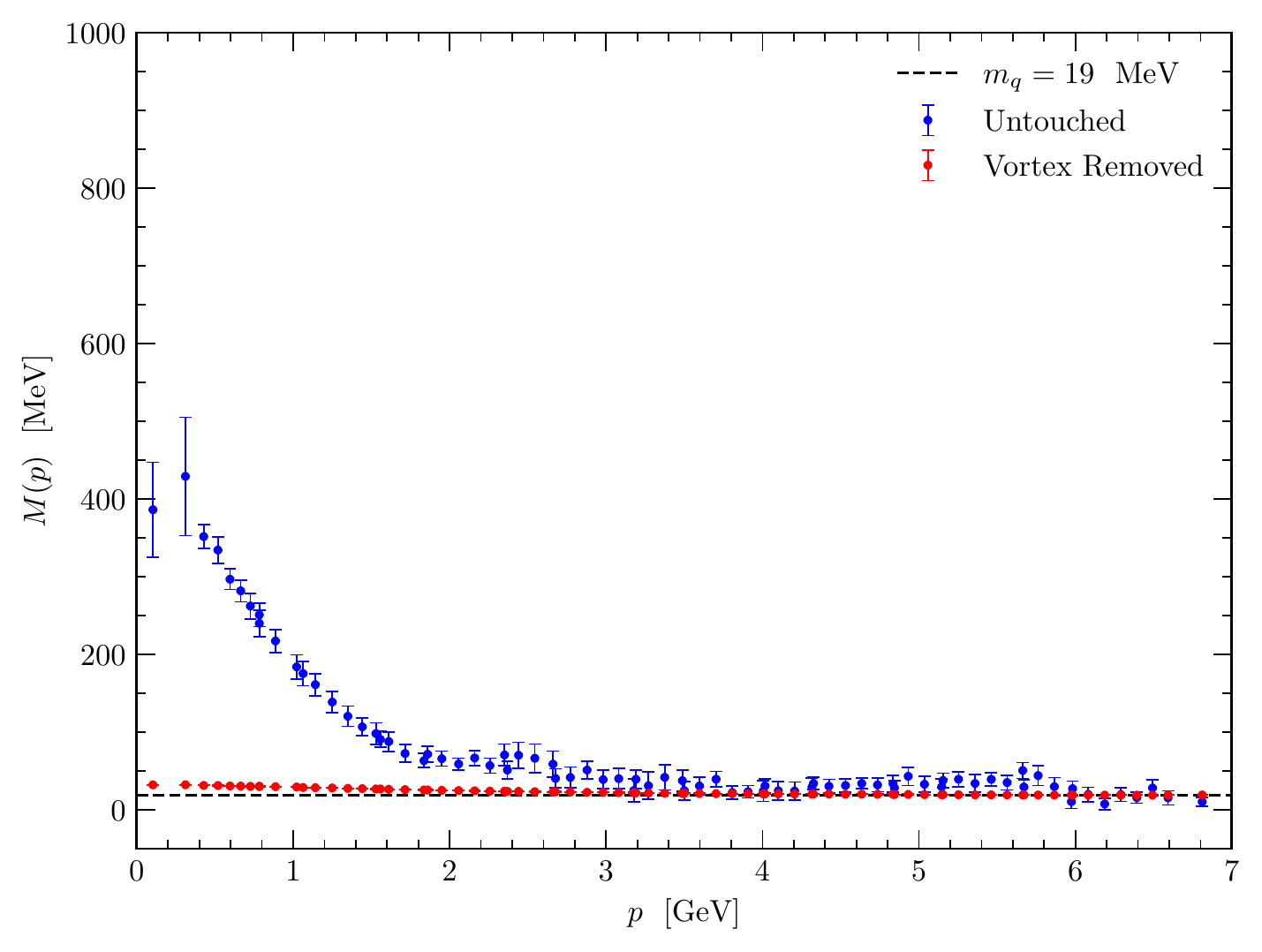}
    \caption{The mass functions, $M(p)$, on untouched and vortex-removed dynamical-fermion
      ensembles for valence-quark masses $m_q=84$ MeV (left), and $m_q=19$ MeV (right). Explicit
      chiral symmetry breaking through the quark mass leaves a remnant of dynamical mass
      generation. 
}
    \label{fig:QCDUTVR}
\end{figure}

Figure~\ref{fig:QCDUTVR} compares the $q^2$-dependent overlap-Dirac quark-propagator Landau-gauge
mass function on the light dynamical-fermion ensemble for two different valence-quark masses.
In comparison to the pure-gauge sector dynamical mass generation is suppressed under
vortex removal in full QCD~\cite{Virgili:2021jnx}.  

While explicit chiral symmetry breaking through the quark mass leaves a
remnant of dynamical mass generation, it is anticipated that for sufficiently light current quark
masses, chiral symmetry will be restored~\cite{Trewartha:2017ive} and dynamical mass generation
will be completely eliminated in the vortex-removed theory.

\section{Summary}

In summary, centre-vortex structure is complex.  Each ground-state configuration is dominated by a
long-distance percolating centre-vortex structure.  In SU(3) gauge field theory, a proliferation
of branching points is observed, with further enhancement as light dynamical fermion degrees of
freedom are introduced in simulating QCD.  There is an approximate doubling in the number of
nontrivial centre charges in the percolating vortex structure as one goes from the pure-gauge
theory to full QCD.  Increased complexity in the vortex paths is also observed as the number of
branching points is significantly increased with the introduction of dynamical fermions.  In short,
dynamical-fermion degrees of freedom radically alter the centre-vortex structure of the
ground-state vacuum fields.  This change in structure acts to improve the phenomenology of centre
vortices better reproducing the string tension, dynamical mass generation and better removing
nonperturbative physics under vortex removal.  This represents a significant advance in the ability
of centre vortices to capture the salient nonperturbative features of QCD.

\section*{Acknowledgements}

DBL thanks Jeff Greensite for several interesting conversations on
centre vortices during the conference.
We thank the PACS-CS Collaboration for making their 2 +1 flavour configurations
available via the International Lattice Data Grid (ILDG). This research was
undertaken with the assistance of resources from the National Computational
Infrastructure (NCI), provided through the National Computational Merit
Allocation Scheme and supported by the Australian Government through Grant No.
LE190100021 via the University of Adelaide Partner Share. This research is
supported by Australian Research Council through Grants No. DP190102215 and
DP210103706. WK is supported by the Pawsey Supercomputing Centre through the
Pawsey Centre for Extreme Scale Readiness (PaCER) program.

%


\bibliography{refs}

\end{document}